\documentclass{PoS}
\usepackage[utf8]{inputenc}

\usepackage{amsmath,dsfont}

\title{Study of $\theta$-Vacua in the 2-d $O(3)$ Model}
\ShortTitle{Study of $\theta$-Vacua in the 2-d $O(3)$ Model}
\author{\speaker{Michael Bögli}, Ferenc Niedermayer, Uwe-Jens Wiese, \\
{Albert Einstein Center for Fundamental Physics \\
Institute for Theoretical Physics, Bern University\\
Sidlerstr.\ 5, 3012 Bern, Switzerland.} \\
        E-mail: \email{boegli@itp.unibe.ch}, \email{niederma@itp.unibe.ch}, \email{wiese@itp.unibe.ch}}

\author{Michele Pepe\\
       {INFN, Sezione di Milano-Bicocca\\ 
Edificio U2, Piazza della Scienza 3, 20126 Milano, Italy.}\\
       E-mail: \email{michele.pepe@mib.infn.it}}

\abstract{ 
We investigate the continuum limit of the step scaling function $\sigma(s, u)$ in the 2-d $O(3)$ model with different $\theta$-vacua. Since we find a different continuum value of $\sigma(s, u)$ for each value of $\theta$, we can conclude that $\theta$ indeed is a relevant parameter of the theory and does not get renormalized non-perturbatively. Furthermore, we confirm the result of the conjectured exact S-matrix theory, which predicts the continuum value at $\theta = \pi$. To obtain high precision data, we use a modified Hasenbusch improved estimator and an action with an optimized constraint, which has very small cut-off effects. The optimized constraint action combines the standard action of the 2-d $O(3)$ model with a topological action. The topological action constrains the angle between neighboring spins and is therefore invariant against small deformations of the field. 
}

\FullConference{The 30 International Symposium on Lattice Field Theory - Lattice 2012,\\
    June 24-29, 2012\\
    Cairns, Australia}

\begin{document}

\section{Introduction}
This is a talk based on \cite{Bogli:2011aa}. The action of the 2-d $O(3)$ model in the continuum is defined as 
\begin{equation}
 S[\vec e] = \frac 1{2g^2} \int \text{d}^2 x\ \partial_\mu \vec e \cdot \partial_\mu \vec e,
 \label{eq:ac_continuum}
\end{equation}
where $g^2$ is the coupling and the unit-vector $\vec e(x) \in S^2$ is the field variable. The 2-d $O(3)$ model shares many features with QCD: Both have important properties including asymptotic freedom and a non-perturbatively generated massgap. One can define a topological charge \cite{Ber81}
\begin{equation}
Q[\vec e] = \frac 1{8\pi} \int \text d^2 x\ \varepsilon_{\mu\nu} \vec e\cdot (\partial_\mu\vec e \times \partial_\nu \vec e),
\end{equation}
which is an integer number and describes how many times the sphere $S^2$ gets covered by $\vec e(x)$, when $x$ runs over a periodic space-time.  There are instantons, which are field configurations with a non-zero topological charge $Q[\vec e]$, which minimize the action $S[\vec e]$. 
%With the topological charge $Q[\vec e]$ 
Because of the presence of non-trivial
topological sectors we can add the $\theta$-vacuum term $i \theta Q[\vec e]$ to the action. Here, we show that the parameter $\theta$ is indeed a relevant parameter in the action, which means it does not get renormalized non-perturbatively. We will show that at each value of $\theta$ there is a different continuum theory.

Haldane showed in the large spin $S$ limit that the low-energy effective field theory of a (1+1)-d  spin chain is the 2-d $O(3)$ model. He further conjectured that integer spin chains correspond to a vacuum angle $\theta = 0$, while half-integer spin chains correspond to $\theta = \pi$ \cite{Hal83}. This conjecture has been confirmed numerically for $\theta = 0$ in \cite{Bot83} and within statistical errors also for $\theta = \pi$ \cite{Bie95}. For spin chains it is well known that for half integer spins the mass gap vanishes \cite{Bet31}. 

The 2-d $O(3)$ model has been studied analytically by the exact S-matrix theory. For $\theta=0$, this theory predicts a mass gap $m(\theta = 0) = \tfrac 8e \Lambda_{\overline{MS}}$, which is generated non-perturbatively and has been verified by lattice simulations. Furthermore it is conjectured that the low-energy effective field theory of the 2-d $O(3)$ model at $\theta = \pi$ is the WZNW model \cite{Wit84}. The exact S-matrix theory even solves analytically the 2-d $O(3)$ model also in the finite volume and provides a prediction for  the step scaling function \cite{Lue91}. Therefore, it is not necessary to take the thermodynamic limit to compare analytic and numerical results. The step scaling function $\sigma(s, u_0)$ describes the scaling behavior of the renormalized coupling $u_0 = m(L)L$ in the finite volume with spacial extend $L$. Here, we will use different lattice actions to confirm the exact S-matrix results. Besides the standard action, we use a topological action \cite{Bie10}, which constrains the maximal angle between neighboring spins. To reduce the cut-off effects, we will combine these two actions in the new ``constraint action``, which has also been studied intensively in \cite{Balog:2012db}.
 In this paper, we also use a variant of a method developed by Hasenbusch \cite{Has95,Bal10} to 
simulate $\theta$-vacuum effects in the 2-d $O(3)$ model with unprecedented per 
mill level precision. For the first time, this numerically confirms the 
conjectured exact S-matrix of the 2-d $O(3)$ model at $\theta = \pi$ 
\cite{Zam86} beyond any reasonable doubt, which also implies that the model 
indeed reduces to the WZNW model at low energies.

%% dislocations
On the lattice, field configurations with non-zero topological charge that minimize the action are known as dislocations. When the dislocation action is less than a critical value, a semi-classical argument suggests that the topological susceptibility $\chi_t = \langle Q^2 \rangle/V$ should suffer from an ultra-violet power-law divergence in the continuum limit \cite{Lue82}.
% Depending on the action one may expect a power-law divergence of this topological susceptibility, which would spoil the continuum limit. 
As we will see later, for each value of $\theta$ the dislocation problem does not prevent $\theta$ to be physically relevant, which suggests that $\chi_t$ only diverges logarithmically.

\section{Lattice Setup}
\subsection{The Lattice Actions}
The simplest lattice discretization of the action (\ref{eq:ac_continuum}) leads us to
\begin{equation}
 S_\text{standard}[\vec e] = -\frac 1{g^2} \sum_{\langle xy\rangle}\vec e_x \cdot \vec e_y,
  \label{eq:ac_std}
\end{equation}
where $\vec e_x \in S^2$ is the field variable, which is now defined on every site $x$ on the lattice. Besides this standard action we also consider the topological action introduced in \cite{Bie10}
\begin{equation}
 S_\text{topological}[\vec e] = \sum_{\langle xy\rangle} s(\vec e_x, \vec e_y), \qquad s(\vec e_x, \vec e_y) = \left\{\begin{array}{ll} 0 \quad  \ & \text{for }\,  \vec e_x \cdot \vec e_y > \cos \delta \\ \infty & \text{else} \end{array} \right..
 \label{eq:ac_top}
\end{equation}
Here, the maximally allowed angle $\delta$ plays the role of the coupling. The topological action only allows configurations where the angle between the field variables on neighboring sites $x$ and $y$ is smaller than $\delta$ ($\vec e_x \cdot \vec e_y > \cos\delta$). Otherwise the action is infinite, which means that the corresponding configurations are not allowed. All allowed configurations have the same action $S_\text{topological}[\vec e] = 0$. As a consequence, this lattice model does not have the correct classical continuum 
limit, it violates the Schwarz inequality between action and topological charge,
and it cannot be treated in perturbation theory. Despite these various 
deficiencies this action still has the correct quantum continuum limit \cite{Bie10}. 

As it will be shown later, the standard action approaches the continuum limit of the step scaling function from above while the topological action approaches it from below. Therefore we combine these two actions in order to reduce the cut-off effects. This we do with the an optimized constraint action
\begin{equation}
 S_\text{constraint}[\vec e] = \sum_{\langle xy\rangle} s'(\vec e_x, \vec e_y), \qquad s'(\vec e_x, \vec e_y) = \left\{\begin{array}{ll} -\frac 1{g^2}\ \vec e_x\cdot \vec e_y \quad  \ &\text{for }\,  \vec e_x \cdot \vec e_y > \cos \delta \\ \infty & \text{else} \end{array} \right..
\end{equation}
Here, $g^2$ is again the coupling, while $\delta$ is a fixed parameter that is tuned to a value which minimizes the cut-off effects. For $\delta = \pi$ this optimized constraint action reduces to the standard action (\ref{eq:ac_std}). On the other hand, if we send $g^2 \rightarrow \infty$, we obtain the topological action (\ref{eq:ac_top}).

To implement the topological charge $Q[\vec e]$ in a discretized form, we triangulate the lattice as shown in figure \ref{fig:triangle}. Then each triangle $\langle xyz\rangle$ gets mapped to an oriented area $A_{\langle xyz\rangle} = 4\pi q_{\langle xyz\rangle}$ on the sphere $S^2$ defined by $\vec e_x,\ \vec e_y$, and $\vec e_z$. If we sum up the areas $q_{\langle xyz\rangle}$ of all triangles, taking in account their orientation, we obtain the topological charge 
\begin{equation}
Q[\vec e] = \sum_{\langle xyz\rangle} q_{\langle xyz\rangle}.
\end{equation}
Using this definition, the topological charge is an integer number, as long as we sum over all triangles in a periodic lattice. The area of one single triangle ${\langle xyz \rangle}$ can be calculated as 
\begin{equation}
 R \text e^{2 \pi i q_{\langle xyz \rangle}}=1 + \vec e_x \cdot \vec e_y +
\vec e_y \cdot \vec e_z + \vec e_z \cdot \vec e_x
% \nonumber \\ &+&
+ i \vec e_x \cdot (\vec e_y \times \vec e_z), \quad R \geq 0.
\end{equation}
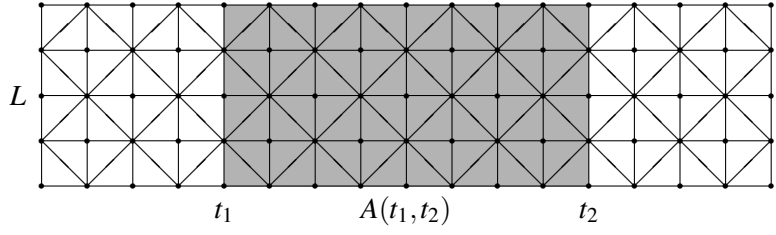
\begin{figure}[htb]
\centering
\setlength{\unitlength}{0.6cm}
\begin{picture}(17.5,5.5)(-1,-1)
\definecolor{mygray}{rgb}{0.7,0.7,0.7}
\put(4,0){\color{mygray}\rule{4.8cm}{2.4cm}}
\multiput(0,0)(0,1){5}{\line(1,0){16}} 
\multiput(0,0)(1,0){17}{\line(0,1){4}} 
\multiput(1,0)(2,0){8}{\line(1,1){1}} 
\multiput(0,1)(2,0){8}{\line(1,-1){1}} 
\multiput(0,1)(2,0){8}{\line(1,1){1}} 
\multiput(1,2)(2,0){8}{\line(1,-1){1}} 
\multiput(1,2)(2,0){8}{\line(1,1){1}} 
\multiput(0,3)(2,0){8}{\line(1,-1){1}} 
\multiput(0,3)(2,0){8}{\line(1,1){1}} 
\multiput(1,4)(2,0){8}{\line(1,-1){1}} 
\multiput(0,0)(0,1){5}{\multiput(0,0)(1,0){17}{\circle*{0.11}}}

% \put(-1,-1){\vector(1,0){3}}
% \put(-1,-1){\vector(0,1){3}}
% \put(2,-1.7){$t$}
% \put(-1.7,2){$x$}
\put(-0.7,1.8){$L$}
\put(3.8,-0.7){$t_1$}
\put(11.8,-0.7){$t_2$}
\put(7.,-0.7){$A(t_1,t_2)$}

 \end{picture}
 \caption{\it Triangulated square lattice: the triangles $\langle xyz \rangle$ 
in the shaded area $A(t_1,t_2)$ carry the topological term 
$i \theta q_{\langle xyz \rangle}$.}
\label{fig:triangle}
\end{figure}

As before, we can add the $\theta$-vacuum term $i\theta Q[\vec e]$ to the action, where $Q[\vec e]$ is now the discretized topological charge.

\subsection{Observables}
To compare results with the exact S-matrix theory, we calculate the step scaling function $\sigma(s, u_0)$ \cite{Lue91}, where $s$ is a rescaling factor. Starting on a volume with spacial extent $L$, one adjusts the coupling $g^2$ in order to obtain the renormalized coupling $u_0 = m(L)L$, where $m(L)$ is the inverse of the finite volume correlation length $m(L) = 1/\xi(L)$. Then, one measures the renormalized coupling $u_1 = m(L')L'$ on the scaled volume with spacial extend $L' = s\cdot L$. Setting the rescaling factor $s=2$ means that we measure the quantity 
\begin{equation}
\sigma(2,u) = 2m\left(2L \right)L\Big|_{m(L)L=u_0} 
\label{eq:ssf}
\end{equation}
on a volume which has twice the spacial extent $L' = 2L$ than the original. To measure the finite volume correlation length $\xi(L)$, we calculate the correlation function 
\begin{equation}
C(t_1,t_2;\theta) = \frac{1}{Z(t_1,t_2;\theta)} \left(\prod_x \int_{S^2} \text d\vec e_x \right)\ 
\vec E(t_1) \cdot \vec E(t_2) \times \text e^{- S[\vec e] + i \theta Q(t_1,t_2)} \ \sim\ 
\text e^{- m(\theta,L)(t_2-t_1)},  
\end{equation}
where $Q(t_1,t_2)$ is the (in general non-integer) topological charge between time-slice $t_1$ and $t_2$, $m(\theta, L)$ is the inverse of the finite volume correlation length $m(\theta,L) = 1/\xi(\theta, L)$, $\vec E(t)$ is the average spin in timeslice $t$ and $Z(t_1,t_2;\theta)$ is the partition function
\begin{equation}
\vec E(t) = \frac 1{N_x} \sum_x \vec e_{(x,t)}, \qquad Z(t_1,t_2;\theta) =  \left(\prod_x \int_{S^2} \text d\vec e_x \right)\ \exp\left(- S[\vec e] + i \theta Q(t_1,t_2)\right).
\end{equation}
To sample Monte Carlo configurations with a positive weight, we need to absorb the phase factor in the observable. This reweighting is obtained as follows
\begin{eqnarray}
 C(t_1,t_2;\theta) &=& \frac{\left\langle \left(\vec E(t_1) \cdot \vec E(t_2) \right)\ \exp(i \theta Q(t_1,t_2)) \right\rangle_{\theta=0} }{\left\langle \exp(i \theta Q(t_1,t_2)) \right\rangle_{\theta=0} } 
 =\frac{ C(t_1,t_2,0) Z(t_1,t_2,\theta)/Z(0) }{Z(t_1,t_2, \theta)/Z(0)} .
\end{eqnarray}

%% cut-off behavior
The cut-off behavior of the lattice step scaling function $\Sigma(\theta, 2, u_0, a/L)$ is known for $\theta = 0$\cite{Bal09}:
\begin{equation}
\Sigma(\theta=0,2,u_0,a/L) = \sigma(\theta=0,2,u_0) + 
\frac{a^2}{L^2} \left[B \log^3(L/a) + C \log^2(L/a) + \dots\right],
\label{eq:cut-off}
\end{equation}
where $B$ and $C$ are fitting parameters and $L/a$ is the number of lattice points in the spatial direction. 

%% hasenbusch method
To obtain high accuracy data, we use an improved estimator first proposed by Hasenbusch \cite{Has95}, which was further developed in \cite{Bal10}. Instead of measuring the estimator $\left(\vec E(t_1) \cdot \vec E(t_2) \right)\ \text e^{i \theta Q(t_1,t_2)}$, we integrate out a certain rotation symmetry. This can be done by rotating all spins after timeslice $t$ with a rotation matrix $X(t):\ \vec e_{(x,t')} \longmapsto X(t) \vec e_{(x,t')}$ for all $t' > t$. Using open boundary conditions in the temporal direction, the action $S[\vec e]$ and the topological charge $Q[\vec e]$ only change in this time slice $t$. For each timeslice $t_1<t\leq t_2$ we now choose four different rotation matrices $X_0(t) = \mathds 1,\ X_1(t),\ X_2(t),\ X_3(t)$, which all do not change the action $S[\vec e]$ (for the construction of $X_i(t)$ see \cite{Bal10}). So in each timeslice $t_1<t\leq t_2$, we insert a matrix 
\begin{equation}
\tilde X(t) = \sum_i X_i(t)\ \exp\left(i \theta Q\left(t,t+1,X_i\left(t\right)\right)\right),
\end{equation}
where $Q(t,t+1,X_i(t))$ is the topological charge after all the spins $\vec e_{(x,t')}$ for $t' > t$ have been rotated with the rotation matrix $X_i(t)$.  With this, we can write improved estimators as
\begin{eqnarray}
 \left(\vec E(t_1) \cdot \vec E(t_2) \right)\ \exp(i \theta Q(t_1,t_2)) &\sim& \vec E(t_1)\ \tilde X(t_1+1)\ \tilde X(t_1+2) \cdots \tilde X(t_2)\ \vec E(t_2) \\
\exp(i \theta Q(t_1,t_2)) &\sim& \Big(\sum_i \text e^{i \theta Q(t_1,t_1+1,X_i(t_1+1))} \Big)\cdots \Big(\sum_i \text e^{i \theta Q(t_2-1,t_2,X_i(t_2))} \Big) \nonumber
\end{eqnarray}

\vspace{-5mm}

\section{Results}
\begin{figure}[htb]
 \centering
 \includegraphics[width=0.54\textwidth]{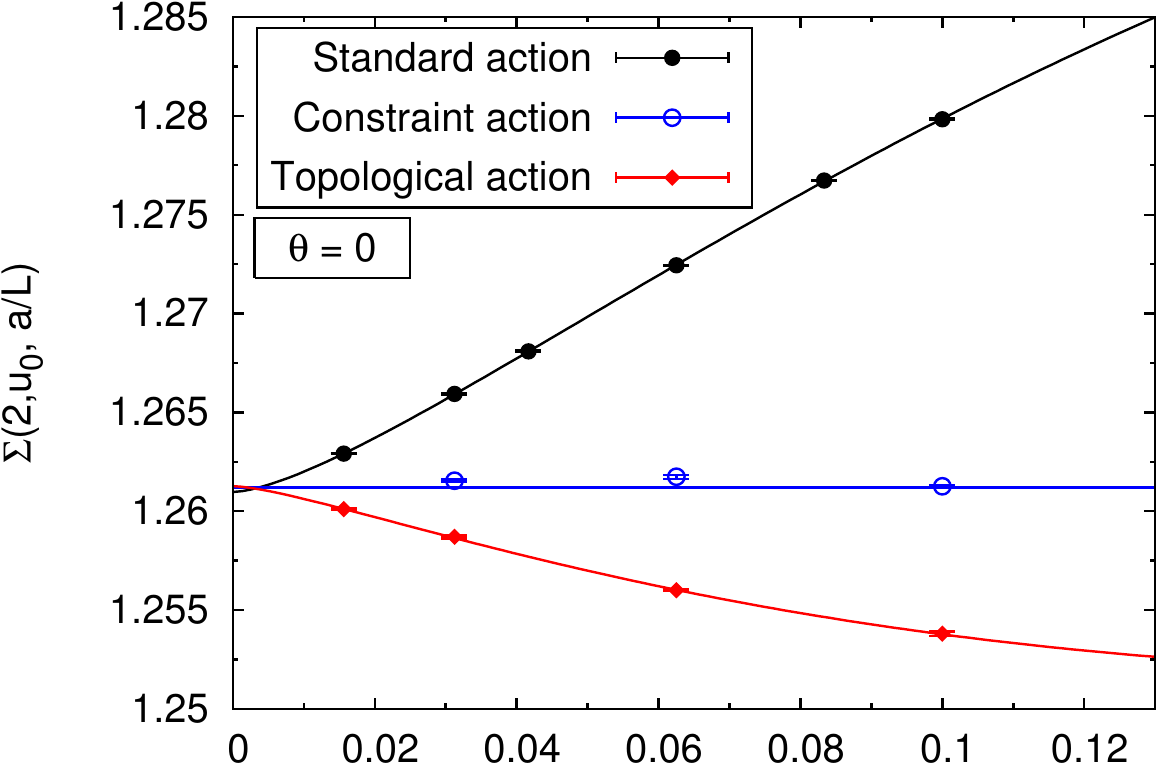}
 \includegraphics[width=0.54\textwidth]{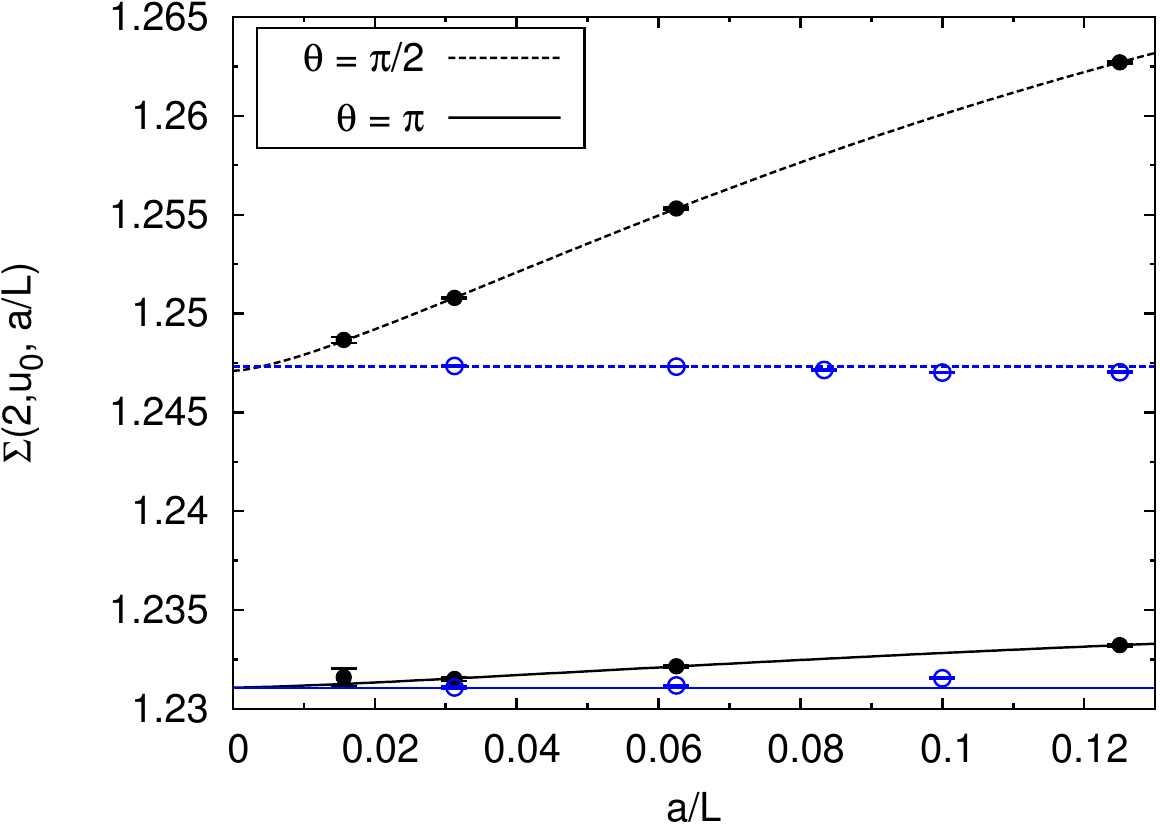}
 \caption{
 Cut-off dependence of the step scaling function 
$\Sigma(\theta,2,u_0 = 1.0595,a/L)$ for three different lattice 
actions: the standard action, the topological lattice action of \cite{Bie10},
and the optimized constraint action with $\cos\delta = - 0.345$, as well as for 
three different values of $\theta = 0$ (top) and $\theta = \frac{\pi}{2}, \pi$ (bottom). The lines are fits based on eq.
%(\ref{eq:cut-off}). 
The horizontal lines represent the analytic continuum results for $\theta = 0$ \cite{Bal10a} and $\theta = \pi$ \cite{Bal11}, and the fitted continuum value for $\theta = \frac{\pi}{2}$.}
 \label{fig:ssf}
\end{figure}

By measuring the step scaling function on different lattice sizes $L/a$, we can plot the cut-off behavior of the different lattice actions as it is done in figure \ref{fig:ssf}. The curves for the standard action and for the topological action are fits of the function (\ref{eq:cut-off}), where $\sigma,\ A$ and $B$ are fitting parameters. 

For $\theta = 0$ the continuum limit (horizontal line) is predicted by the exact S-matrix theory to be at $\sigma(0, 2, 1.0595) =  1.2612103$ \cite{Bal10a}. As one can see, the standard action approaches this continuum limit from above, while the topological action approaches the continuum limit from below. The maximum angle $\delta$ of the constraint action is tuned to the continuum value at a lattice size $L/a = 10$ and takes an optimal value at $\cos(\delta) = -0.345$. Even for finer lattices the cut-off effects of the constraint action are remarkably small and only at the per mill level.

At $\theta = \pi$ the exact S-matrix theory again predicts a continuum result (horizontal line) of $\sigma(\pi,2,1.0595) = 1.231064$ \cite{Bal11}. This result is confirmed by extrapolating the standard action or the constraint action. For the constraint action, we still use the maximal angle $\delta$, which has been tuned at $\theta = 0$. Also in this case the cut-off effects of the constraint action are extremely small.

For $\theta = \pi/2$, we do not have an analytic prediction of the continuum limit and we do not know whether the function (\ref{eq:cut-off}) is still applicable to describe the cut-off effects. Nevertheless, we fit this function to the cut-off effects of the standard action, which gives a small $\chi^2/$d.o.f. Extrapolating to the continuum limit agrees with an estimator of the continuum limit for the constraint action (horizontal line). 
Again, the constraint action has a maximum angle constraint $\delta$, which has been tuned at $\theta = 0$, but still shows an extremely good cut-off behavior.

\begin{figure}[htb]
 \centering
 \includegraphics[width=0.55\textwidth]{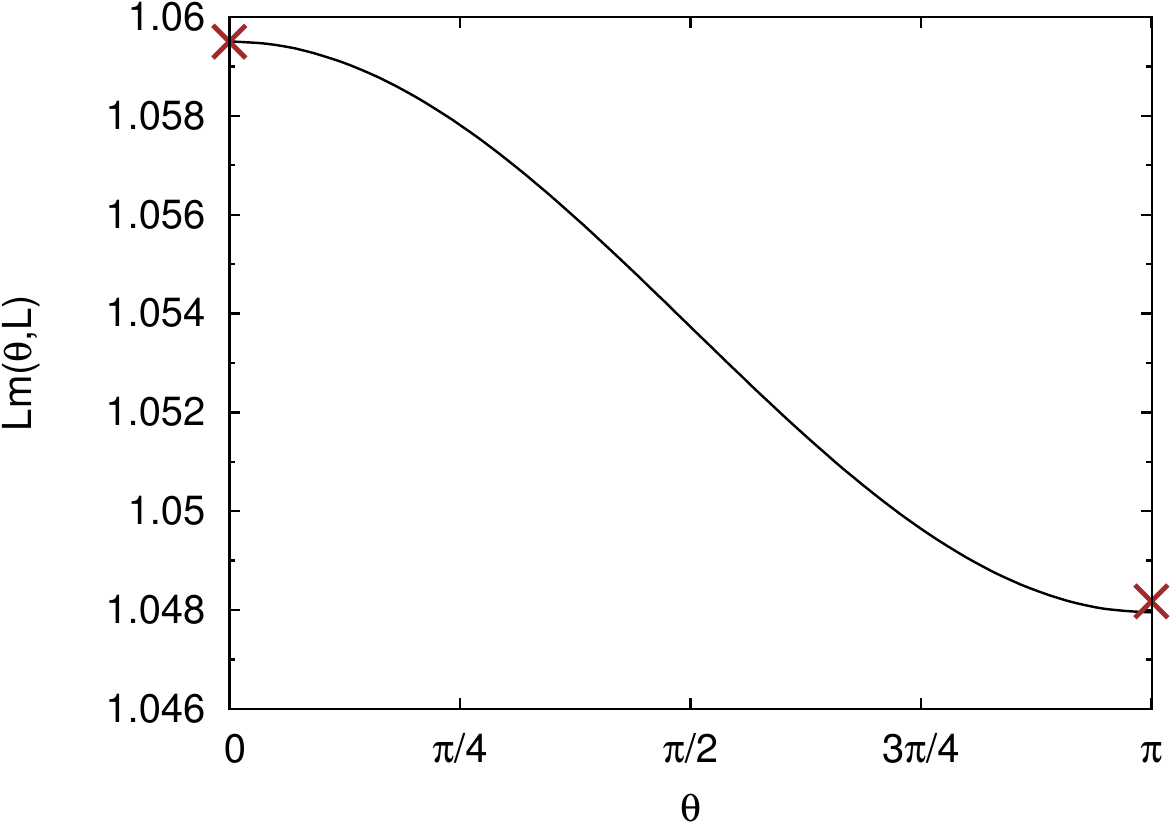}
 \caption{The $\theta$-dependent massgap $L m(\theta,L)$ at 
$L m(0,L) = 1.0595$ using the optimized constraint action for $L = 24 a$, 
compared to the analytic result at $\theta = \pi$ \cite{Bal11} (cross).}
 \label{fig:zerotopi}
\end{figure}

Figure \ref{fig:ssf} also shows that all continuum limits (at $\theta = 0,\pi/2, \pi$) are significantly different, which means that each value of $\theta$ indeed corresponds to a different theory. Hence $\theta$ is a relevant parameter that does not renormalize to 0 or $\pi$ non-perturbatively, as one might have expected due to the presence of dislocations. This can also be seen in figure \ref{fig:zerotopi}, where we plot $m(\theta, L)L$ as a function of $\theta$, keeping the coupling fixed. For $\theta = 0$, we fix the value to $m(0, L)L = 1.0595$, at $\theta = \pi$ one obtains $m(\pi, L)L = 1.048175$ \cite{Bal11}.

\newpage
\section{Conclusion}

Using an efficient modification of Hasenbusch's improved estimator allowed us to show that the $\theta$ parameter in the 2-d $O(3)$ model is indeed a relevant parameter and does not get renormalized non-perturbatively. We found a different continuum limit for each value of $\theta$ (here shown for $\theta = 0, \pi/2, \pi$). Dislocations do not spoil this continuum limit even for $\theta \neq 0$. We confirmed the exact S-matrix conjecture for the step scaling function at $\theta = \pi$, which also implies that the model indeed reduces to the WZNW model at low energies. 

In the simulations we used different lattice actions. The constraint action combines the standard action with a topological action. The topological action restricts the maximal angle between neighboring spins and is thus invariant under small deformations of the field. In the constraint action this maximal angle is tuned in order to reduce the cut-off effects. Our results show that only a single tuning is necessary (at $L/a = 10,\ \theta = 0$) to once and for all fix this maximal angle. Other lattice sizes or other values of $\theta$ have extremely small cut-off effects of the step scaling function which all lie in the per mill level or beyond.

We are indebted to J.\ Balog for providing exact values of the step scaling 
function at $\theta = \pi$ prior to publication. This work has been supported 
by the Regione Lombardia and CILEA Consortium through a LISA 2011 grant, as 
well as by the Schweizerischer Nationalfonds (SNF).

\end{document}